\documentstyle[12pt]{article}
\newcommand\be{\begin{equation}}
\newcommand\ee{\end{equation}}
\newcommand\bea{\begin{eqnarray}}
\newcommand\eea{\end{eqnarray}}
\newcommand\ket[1]{|#1\rangle}

\newcommand\braket[2]{\langle #1|#2\rangle}
\newcommand{\fatalpha}{{\bf \alpha \kern -0.44em \alpha}}
\newcommand{\fatsigma}{{\bf \sigma \kern -0.54em \sigma}}
\newcommand{\tpchi}{{\bf \chi \kern -0.35em \chi}}
\newcommand{\llambda}{{\bf \lambda \kern -0.45em \lambda}}

\renewcommand{\theequation}{\arabic{equation}}
\renewcommand{\theequation}{\thesection-\arabic{equation}}
\bibliography{plain}
\title{\bf Perfect state transfer over distance-regular spin networks}\vspace{20mm}
\author{ M. A. Jafarizadeh$^{a,b,c}$
 \thanks{E-mail:jafarizadeh@tabrizu.ac.ir} and
 R. Sufiani$^{a,b}$
 \thanks{E-mail:sofiani@tabrizu.ac.ir}
 \\ $^a${\small Department of Theoretical Physics and Astrophysics,
University of Tabriz, Tabriz 51664, Iran.} \\ $^b${\small
Institute for Studies in Theoretical Physics and Mathematics,
Tehran 19395-1795, Iran.} \\ $^c${\small Research Institute for
Fundamental Sciences, Tabriz 51664, Iran.}} \pagebreak

\vspace{20mm}
\begin{document}
\maketitle \vspace{15mm}
\newpage
\begin{abstract}
By considering distance-regular graphs as spin networks, first we
introduce some particular spin Hamiltonians which are extended
version of those of Refs.\cite{8,9''}. Then, by using spectral
analysis techniques and algebraic combinatoric structure of
distance-regular graphs such as stratification introduced in
\cite{obata, js} and Bose-Mesner algebra, we give a method for
finding a set of coupling constants in the Hamiltonians so that a
particular state initially encoded on one site of a network will
evolve freely to the opposite site without any dynamical controls,
i.e., we show that how to derive the parameters of the system so
that perfect state transfer (PST) can be achieved. As examples,
the cycle networks with even number of vertices and
$d$-dimensional hypercube networks are considered in details and
the method is applied for some important distance-regular networks
in
 appendix.\\


{\bf PACs Index: 01.55.+b, 02.10.Yn }
\end{abstract}

\vspace{70mm}
\newpage
\section{Introduction}
The transfer of a quantum state from one part of a physical unit,
e.g., a qubit, to another part is a crucial ingredient for many
quantum information processing protocols \cite{1}. Currently,
there are several ways of moving data around in a quantum
computer. While some methods transfer quantum states by moving
them down a linear array of qubits, there are others which exploit
the quantum property of entanglement for teleporting quantum
states between distant qubits \cite{2,3}. In a
quantum-communication scenario, the transfer of quantum states
from one location $A$ to another location $B$, is rather explicit,
since the goal is the communication between distant parties $A$
and $B$ ( e.g., by means of photon transmission). Equally, in the
interior of quantum computers good communication between different
parts of the system is essential. The need is thus to transfer
quantum states and generate entanglement between different regions
contained within the system. There are various physical systems
that can serve as quantum channels, one of them being a quantum
spin system. This can be generally defined as a collection of
interacting qubits (spin-1/2 particles) on a graph, whose dynamics
is governed by a suitable Hamiltonian, e.g., the Heisenberg or
$XY$ Hamiltonian.

Quantum communication over short distances through a spin chain,
in which adjacent qubits are coupled by equal strength has been
studied in detail, and an expression for the fidelity of quantum
state transfer has been obtained \cite{4,5}. Similarly, in Ref.
\cite{6}, near perfect state transfer was achieved for uniform
couplings provided a spatially varying magnetic field was
introduced. The propagation of quantum information in rings has
been also investigated in \cite{7}. In our work we focus on the
situation in which state transference is perfect, i.e., the
fidelity is unity, and in which we can design spin networks such
that this can be achieved over arbitrarily long distances. We will
also consider the case in which no external control is required
during the state transference, i.e., we consider the case in which
we have, after manufacturing the network, no further control over
its dynamics. In general this will lead us to think about more
complicated spin networks than the linear chain or chains with
preengineered nearest-neighbor interaction strengths. We provide
two alternative methods for understanding how perfect state
transfer is achieved with preengineered couplings. This paper
expands and extends the work done in \cite{8,9''}.
 We will consider distance-regular graphs as spin networks in the
 sense that with each vertex of a distance-regular graph a qubit
 or a spin is associated (although qubits represent generic
two state systems, for convenience of exposition we will use the
term spin as it provides a simple physical picture of the
network). Then, duo to the fact that distance-regular graphs are
underlying graphs of association schemes (see for example
\cite{Ass.sch., jss}), we use their algebraic properties in order
to find suitable coupling constants in some particular spin
Hamiltonians so that perfect transference of a quantum state
between antipodes of the networks can be achieved. More clearly,
for a given distance-regular network first we stratify the network
with respect to an arbitrary chosen vertex of the network called
reference vertex (for details about stratification of graphs, see
\cite{obata, js, jss}). Then, we consider coupling constants so
that vertices belonging to the same stratum with respect to the
reference vertex possess the same coupling strength with the
reference vertex whereas vertices belonging to distinct strata
possess different coupling strengths. Then we give a method for
finding a suitable set of coupling constants so that PST over
antipodes of the networks be possible. As examples we will
consider the cycle networks with even number of vertices and $d$-
dimensional hypercube networks in details and some important
distance-regular networks in an appendix.

The organization of the paper is as follows: In section 2, we
review some preliminary facts about association schemes,
stratification, distance-regular graphs and spectral analysis
techniques. Section $3$ is devoted to perfect state transfer (PST)
over antipodes of distance-regular networks, where a method for
finding suitable coupling constants in particular spin
Hamiltonians so that PST be possible, is given. The paper is ended
with a brief conclusion and an appendix.
\section{Preliminaries}
In this section we will first review some preliminary facts about
distance-regular graphs, corresponding stratification and spectral
distribution associated with graphs.
\subsection{Association schemes}
First we recall the definition of association schemes. For further
information on association schemes, the reader is
referred to Ref. \cite{Ass.sch.}.\\
\textbf{Definition.} An association scheme with $d$ associate
classes on a finite set $V$ is a set of matrices $A_0, A_1, ...,
A_d$ in $R^{V\times V}$, all of whose entries are equal to $0$ or
$1$, such that\\
(i) $A_0 = I_v$;\\
(ii) $A_i$ is symmetric for $i=1,..., d$;\\
(iii) for all $i, j$ in $\{0,1,...,d\}$, the product $A_iA_j$ is a
linear combination of $A_0, A_1,...,A_d$;\\
(iv) none of the $A_i$ is equal to $O_v$, and $\sum_{i=0}^dA_i=J_v$,
where  $v:=|V|$ and $J_v$ is an $v\times v$ all one matrix.\\
It should be noticed that, since $A_i$ is symmetric with entries in
$\{0,1\}$, the diagonal entries of $A^2_i$ are the row-sums of
$A_i$. Condition (iii) implies that $A^2_i$ has a constant element,
say $\kappa_i$, on its diagonal. Therefore every row and every
column of $A_i$ contains $\kappa_i$ entries equal to $1$. Hence
$A_iJ_v=J_vA_i=\kappa_iJ_v$. Moreover, $A_0A_i=A_iA_0=A_i$.

From condition (iii), one can write
\begin{equation}\label{ss}
A_iA_j=\sum_{k=0}^{d}p_{ij}^kA_{k},
\end{equation}
which implies that the adjacency matrices $A_0, A_1, ..., A_d$ form
a basis for a commutative algebra \textsf{A} known as Bose-Mesner
algebra associated with the association scheme. This algebra has a
second basis $E_0, ...,E_d$ such that, $E_iE_j= \delta_{ij}E_i$ and
$\sum_{i=0}^d E_i=I$ with $E_0=1/v J_v$ \cite{Ass.sch.}. The
matrices $E_i$ for $0\leq i\leq d$ are known as primitive
idempotents of $Y$. Furthermore, there are matrices $\mathrm{P}$ and
$\mathrm{Q}$ such that the two bases of the Bose-Mesner algebras can
be related to each other as follows
$$A_i=\sum_{j=0}^d{\mathrm{P}}_{ji}E_j,\;\;\ 0\leq j\leq d,$$
\begin{equation}\label{ap}
E_i=\frac{1}{v}\sum_{j=0}^d{\mathrm{Q}}_{ji}A_j,\;\;\ 0\leq j\leq d.
\end{equation}
Then, clearly we have
\begin{equation}\label{pq}
{\mathrm{P}}{\mathrm{Q}}={\mathrm{Q}}{\mathrm{P}}=vI.
\end{equation}
It also follows that
\begin{equation}\label{eign}
A_jE_i={\mathrm{P}}_{ij}E_i,
\end{equation}
which indicates that ${\mathrm{P}}_{ij}$ is the $i$-th eigenvalue of
$A_j$ and that the columns of $E_i$ are corresponding eigenvectors.
Also, $m_i:=tr E_i=v\langle
\alpha|E_i|\alpha\rangle={\mathrm{Q}}_{i0}$ (where, we have used the
fact that $\langle\alpha|E_i|\alpha\rangle$ is independent of the
choice of $\alpha\in V$, see Eq.(2-3)) is the rank of the idempotent
$E_i$ which gives the multiplicity of the eigenvalue
${\mathrm{P}}_{ij}$ of $A_j$ (provided that ${\mathrm{P}}_{ij}\neq
{\mathrm{P}}_{kj}$ for $k \neq i$). Clearly, we have $m_0=1$ and
$\sum_{i=0}^dm_i=v$ since $\sum_{i=0}^dE_i=I$.

The underlying network of an association scheme $\Gamma=(V,E)$ is an
undirected connected network with adjacency matrix $A\equiv A_1$.
Obviously replacing $A_1$ with one of the other adjacency matrices
$A_i$, $i\neq 0,1$ will also gives us an underlying network
$\Gamma'=(V,E')$ (not necessarily a connected network) with the same
set of vertices but a new set of edges.

As we will see in subsection $2.3$, in the case of distance-regular
networks, the adjacency matrices $A_j$ are polynomials of the
adjacency matrix $A\equiv A_1$, i.e., $A_j=P_j(A)$, where $P_j$ is a
polynomial of degree $j$, then the eigenvalues ${\mathrm{P}}_{ij}$
in (\ref{eign}) are polynomials of eigenvalues
${\mathrm{P}}_{i1}\equiv \lambda_i$ (eigenvalues of the adjacency
matrix $A$). This indicates that in distance-regular graphs, the
matrix $\mathrm{P^t}$ is a polynomial transformation \cite{Puschel}
as
\begin{equation}\label{push}
\mathrm{P^t}=\left(\begin{array}{ccccc}
    1 & 1 & \ldots & 1 \\
      P_1(\lambda_0) & P_1(\lambda_1) & \ldots & P_1(\lambda_d) \\
      P_2(\lambda_0) & P_2(\lambda_1) & \ldots & P_2(\lambda_d) \\
      \vdots & \vdots & \ldots & \vdots \\
      P_d(\lambda_0) & P_d(\lambda_1) & \ldots & P_d(\lambda_d) \\
    \end{array}\right)
    \end{equation}
or ${\mathrm{P}}_{ji}=P_i(\lambda_j)$.
\subsection{Stratifications}
For a given vertex $\alpha\in V$, let $\Gamma_i(\alpha):=\{\beta\in
V: (\alpha, \beta)\in R_i\}$  denotes the set of all vertices having
the relation $R_i$ with $\alpha$. Then, the vertex set $V$ can be
written as disjoint union of $\Gamma_i(\alpha)$ for $i=0,1,2,...,d$,
i.e.,
 \begin{equation}\label{asso1}
 V=\bigcup_{i=0}^{d}\Gamma_{i}(\alpha).
 \end{equation}
We fix a point $o\in V$ as an origin of the underlying graph of an
association scheme, called reference vertex. Then, the relation
(\ref{asso1}) stratifies the graph into a disjoint union of
associate classes $\Gamma_{i}(o)$ (called the $i$-th stratum with
respect to $o$). Let $l^2(V)$ denote the Hilbert space of $C$-valued
square-summable functions on $V$. With each associate class
$\Gamma_{i}(o)$ we associate a unit vector in $l^2(V)$ defined by
\begin{equation}\label{unitv}
\ket{\phi_{i}}=\frac{1}{\sqrt{\kappa_i}}\sum_{\alpha\in
\Gamma_{i}(o)}\ket{\alpha},
\end{equation}
where, $\ket{\alpha}$ denotes the eigenket of $\alpha$-th vertex at
the associate class $\Gamma_{i}(o)$ and $\kappa_i=|\Gamma_{i}(o)|$
is called the $i$-th valency of the graph. Now, let $A_i$ be the
adjacency matrix of the graph $\Gamma=(V,R)$. Then, for the
reference state $\ket{\phi_0}$ ($\ket{\phi_0}=\ket{o}$, with $o\in
V$ as reference vertex), one can write
\begin{equation}\label{Foc1}
A_i\ket{\phi_0}=\sum_{\beta\in \Gamma_{i}(o)}\ket{\beta}.
\end{equation}
Then, by using (\ref{unitv}) and (\ref{Foc1}), we obtain
\begin{equation}\label{Foc2}
A_i\ket{\phi_0}=\sqrt{\kappa_i}\ket{\phi_i}.
\end{equation}
One should notice that, in underlying networks of association
schemes, stratification is reference state independent, namely one
can choose any arbitrary vertex as a reference state.
\subsection{Distance-regular networks and spectral techniques}
Distance-regular graphs are underlying graphs of so called
$P$-polynomial association schemes \cite{Ass.sch.}, where the
adjacency matrices $A_i$ are defined based on shortest path
distance. More clearly, if distance between nodes $\alpha,\beta\in
V$ denoted by $\partial(\alpha, \beta)$ be the length of the
shortest walk connecting $\alpha$ and $\beta$ (recall that a finite
sequence $\alpha_0, \alpha_1,..., \alpha_n \in V$ is called a walk
of length $n$ if $\alpha_{k-1}\sim \alpha_k$ for all $k=1, 2,...,
n$, where $\alpha_{k-1}\sim \alpha_k$ means that $\alpha_{k-1}$ is
adjacent with $\alpha_{k}$), then the adjacency matrices $A_i$ for
$i=0,1,...,d$ in distance-regular graphs are defined as:
$(A_i)_{\alpha,\beta}=1$ if and only if $\partial(\alpha, \beta)=i$
and $(A_i)_{\alpha,\beta}=0$ otherwise, where
$d:=$max$\{\partial(\alpha, \beta): \alpha, \beta\in V \}$ is
diameter of the graph.

For distance-regular graphs, the non-zero intersection numbers are
given by
\begin{equation}\label{abc}
 a_i=p_{i1}^i, \;\;\;\  b_i=p_{i+1,1}^i, \;\;\;\
 c_i=p_{i-1,1}^i\;\ ,
\end{equation}
respectively. The intersection numbers (\ref{abc}) and the valencies
$\kappa_i$ with $\kappa_1\equiv\kappa$($=deg(\alpha)$, for each
vertex $\alpha$) satisfy the following obvious conditions
$$a_i+b_i+c_i=\kappa,\;\;\ \kappa_{i-1}b_{i-1}=\kappa_ic_i ,\;\;\
i=1,...,d,$$
\begin{equation}\label{intersec}
\kappa_0=c_1=1,\;\;\;\ b_0=\kappa_1=\kappa, \;\;\;\ (c_0=b_d=0).
\end{equation}
Thus all parameters of a distance-regular graph can be obtained from
its intersection array $\{b_0,...,b_{d-1};c_1,...,c_d\}$. Then, it
can be shown that the following recursion relations are satisfied
$$A_1A_i=b_{i-1}A_{i-1}+a_iA_i+c_{i+1}A_{i+1},\;\
i=1,2,...,d-1,$$
\begin{equation}\label{P0}
A_1A_d=b_{d-1}A_{d-1}+(\kappa-c_d)A_d.
\end{equation}
The recursion relations (\ref{P0}) imply that
\begin{equation}\label{P1}
A_i=P_i(A),\;\ i=0,1,...,d.
\end{equation}
Then, one can easily obtain the following three term recursion
relations for the unit vectors $\ket{\phi_i}$, $i=0,1,...,d$
\begin{equation}\label{trt}
A\ket{\phi_i}=\beta_{i+1}\ket{\phi_{i+1}}+\alpha_i\ket{\phi_i}+\beta_{i}\ket{\phi_{i-1}},
\end{equation}
where, the coefficients $\alpha_i$ and $\beta_i$ are defined as
\begin{equation}\label{omegal}
\alpha_0=0,\;\;\ \alpha_k\equiv a_k=\kappa-b_{k}-c_{k},\;\;\;\;\
\omega_k\equiv\beta^2_k=b_{k-1}c_{k},\;\;\ k=1,...,d,
\end{equation}
(see Ref. \cite{js,jss1,js1} for more details).

Now, we recall some preliminary facts about spectral techniques used
in the paper, where more details have been given in Refs.
\cite{js,jss1,js1}.

For any pair $(A,\ket{\phi_0})$ of a matrix $A$ and a vector
$\ket{\phi_0}$, one can assign a measure $\mu$ as follows
\begin{equation}\label{sp1}
\mu(x)=\braket{ \phi_0}{E(x)|\phi_0},
\end{equation}
 where
$E(x)=\sum_i|u_i\rangle\langle u_i|$ is the operator of projection
onto the eigenspace of $A$ corresponding to eigenvalue $x$, i.e.,
\begin{equation}
A=\int x E(x)dx.
\end{equation}
Then, for any polynomial $P(A)$ we have
\begin{equation}\label{sp2}
P(A)=\int P(x)E(x)dx,
\end{equation}
where for discrete spectrum the above integrals are replaced by
summation. Therefore, using the relations (\ref{sp1}) and
(\ref{sp2}), the expectation value of powers of adjacency matrix $A$
over reference vector $\ket{\phi_0}$ can be written as
\begin{equation}\label{v2}
\braket{\phi_{0}}{A^m|\phi_0}=\int_{R}x^m\mu(dx), \;\;\;\;\
m=0,1,2,....
\end{equation}
Obviously, the relation (\ref{v2}) implies an isomorphism from the
Hilbert space of the stratification onto the closed linear span of
the orthogonal polynomials with respect to the measure $\mu$. More
clearly, from orthonormality of the unit vectors $\ket{\phi_i}$
(with $\ket{\phi_0}$ as unit vector assigned to reference node) we
have
\begin{equation}\label{ortpo}
\delta_{ij}=\langle\phi_i|\phi_j\rangle=\frac{1}{\sqrt{\kappa_i\kappa_j}}\braket{\phi_{0}}{A_iA_j|\phi_0}=\int_{R}P'_i(x)P'_j(x)\mu(dx),
\end{equation}
with $P'_i(A):=\frac{1}{\sqrt{\kappa_i}}P_i(A)$ where, we have used
the equations (\ref{Foc2}) and (\ref{P1}) to write
\begin{equation}\label{xx}
\ket{\phi_i}=\frac{1}{\sqrt{\kappa_i}}A_i\ket{\phi_0}=\frac{1}{\sqrt{\kappa_i}}P_i(A)\ket{\phi_0}\equiv
P'_i(A)\ket{\phi_0}.
\end{equation}
 Now, by substituting (\ref{xx}) in (\ref{trt}) and rescaling $P'_k$
as $Q_k=\beta_1...\beta_kP'_k$, the spectral distribution $\mu$
under question will be characterized by the property of orthonormal
polynomials $\{Q_k\}$ defined recurrently by
$$ Q_0(x)=1, \;\;\;\;\;\
Q_1(x)=x,$$
\begin{equation}\label{op}
xQ_k(x)=Q_{k+1}(x)+\alpha_{k}Q_k(x)+\beta_k^2Q_{k-1}(x),\;\;\ k\geq
1.
\end{equation}
where, the coefficients $\alpha_i$ and $\beta_i$ are defined as
\begin{equation}\label{omegal}
\alpha_k=\kappa-b_{k}-c_{k},\;\;\;\;\
\omega_k\equiv\beta^2_k=b_{k-1}c_{k},\;\;\ k=1,...,d,
\end{equation}
If such a spectral distribution is unique, the spectral
distribution $\mu$ is determined by the identity
\begin{equation}\label{sti}
G_{\mu}(x)=\int_{R}\frac{\mu(dy)}{x-y}=\frac{1}{x-\alpha_0-\frac{\beta_1^2}{x-\alpha_1-\frac{\beta_2^2}
{x-\alpha_2-\frac{\beta_3^2}{x-\alpha_3-\cdots}}}}=\frac{Q_{d}^{(1)}(x)}{Q_{d+1}(x)}=\sum_{l=0}^{d}
\frac{\gamma_l}{x-x_l},
\end{equation}
where, $x_l$ are the roots  of polynomial $Q_{d+1}(x)$.
$G_{\mu}(x)$ is called the Stieltjes/Hilbert transform of spectral
distribution $\mu$ or Stieltjes function and polynomials
$\{Q_{k}^{(1)}\}$ are defined recurrently as
$$Q_{0}^{(1)}(x)=1, \;\;\;\;\;\
    Q_{1}^{(1)}(x)=x-\alpha_1,$$
\begin{equation}\label{oq}
xQ_{k}^{(1)}(x)=Q_{k+1}^{(1)}(x)+\alpha_{k+1}Q_{k}^{(1)}(x)+\beta_{k+1}^2Q_{k-1}^{(1)}(x),\;\;\
k\geq 1,
\end{equation}
respectively. The coefficients $\gamma_l$ appearing in (\ref{sti})
are calculated as
\begin{equation}\label{Gauss}
\gamma_l:=\lim_{x\rightarrow x_l}[(x-x_l)G_{\mu}(x)]
\end{equation}
Now let $G_{\mu}(z)$ is known, then the spectral distribution
$\mu$ can be recovered from $G_{\mu}(z)$ by means of the
Stieltjes/Hilbert inversion formula as
\begin{equation}\label{m1}
\mu(y)-\mu(x)=-\frac{1}{\pi}\lim_{v\longrightarrow
0^+}\int_{x}^{y}Im\{G_{\mu}(u+iv)\}du.
\end{equation}
Substituting the right hand side of (\ref{sti}) in (\ref{m1}), the
spectral distribution can be determined in terms of $x_l,
l=1,2,...$ and  Guass quadrature constants $\gamma_l, l=1,2,... $
as
\begin{equation}\label{m}
\mu=\sum_l \gamma_l\delta(x-x_l)
\end{equation}
(for more details see Refs.\cite{st, obh,tsc,obah}).
\section{Perfect State Transfer (PST) over antipodes of distance-regular networks}
\subsection{State Transfer in
Quantum Spin Systems}
The PST algorithm was proposed by Christandl et al. [1,2], and it
can be implemented in the XY chain. The algorithm can transfer an
arbitrary quantum state between the two ends of the chain in a fixed
period time, only using XY interactions. For one-dimensional
fermionic chains, the model of a system consisting of spinless
fermions (or bosons) hopping freely in a network of $N$ lattice
sites can be mapped to spin chains in which spins are coupled
through the XY Hamiltonian \be H= \frac{1}{2} \sum_{j=1}^{N-1} J_j
(\sigma^x_j\sigma^x_{j+1}+\sigma^y_j\sigma^y_{j+1})+\frac{1}{2}
\sum_{j=1}^{N} \lambda_j (\sigma^z_j+\mathbf{1}), \ee by the
Jordan-Wigner transformation, where $J_j$ is the time-independent
coupling constant between nearest-neighbor sites $j$ and $j+1$, and
$\lambda_j$ represents the strength of the external static potential
at site $j$.

A quantum spin system associated with a simple, connected, finite
graph $G=(V,E)$ as a spin network is defined by attaching a spin-1/2
particle to each vertex of the graph so that to each vertex $i\in V$
one can associate a Hilbert space ${\mathcal{H}}_i\simeq
{\mathcal{C}}^2$. The Hilbert space associated with $G$ is then
given by
\begin{equation}
{\mathcal{H}}_G = \otimes_{_{i\in V}}{\mathcal{H}}_i =
({\mathcal{C}}^2)^{\otimes N},
\end{equation}
 where $N:=|V|$ denotes the total
number of vertices in $G$. On the other hand, quantum state transfer
over a network is similar to the quantum random walk problem, where
a variety of networks are equivalent to one-dimensional chains
[1,22]. Therefore, it can be focused on a chain of $N$ sites. For
$j=1,2, . . . ,N$, let $|j\rangle$ be the state where a single
fermion (or boson) is at the site $j$ but is in the empty state
$|0\rangle$ for all other sites and $|0\rangle$ be the vacuum state
where all sites are empty. For spin chains, $|0\rangle$ corresponds
to the state where all the spins are in the spin-down state
$|\downarrow\rangle$ and $|j\rangle$ corresponds to a spin-up state
$|\uparrow\rangle$ for the $j$th spin and spin-down for all other
spins. The Hamiltonian in this single-particle subspace can be
written in a tridiagonal form, which is real and symmetric: \be
H=\left( \begin{array}{ccccc}
                                            \lambda_1 & J_1 & 0 & \ldots & 0 \\
                                            J_1 & \lambda_2 & J_2 & \ldots & 0 \\
                                            0 & J_2 & \lambda_3 &  \ldots& 0 \\
                                            \vdots & \vdots & \vdots & \ddots & J_{N-1} \\
                                            0 & 0 & 0 & J_{N-1} &
                                            \lambda_N
                                          \end{array}
\right) \ee The quantum state transfer protocol involves two steps:
initialization and evolution. First, a quantum state
$\ket{\psi}_A=\alpha\ket{0}_A+\beta\ket{1}_A\in {\mathcal{H}}_A$
(with $\alpha,\beta\in \mathcal{C}$ and $|\alpha|^2+|\beta|^2=1$) to
be transmitted is created. The state of the entire spin system after
this step is given by
\be\label{eq1}\ket{\psi(t=0)}=\ket{\psi_A0...00_B}=\alpha\ket{0_A0...00_B}+\beta\ket{1_A0...00_B}=\alpha\ket{\b{0}}+\beta\ket{A},\ee
with $\ket{\b{0}}:=\ket{0_A0...00_B}$. Then, the network couplings
are switched on and the whole system is allowed to evolve under
$U(t)=e^{-iHt}$ for a fixed time interval, say $t_0$. The final
state becomes  \be \ket{\psi(t_0)} =
\alpha\ket{\b{0}}+\beta\sum_{j=1}^Nf_{jA}(t_0)\ket{j} \ee where,
$f_{jA}(t_0):=\langle j|e^{-iHt_0}|A\rangle$. Any site $B$ is in a
mixed state if $|f_{AB}(t_0)|<1$, which also implies that the state
transfer from site $A$ to $B$ is imperfect. In this paper, we will
focus only on perfect state transfer. This means that we consider
the condition \be\label{eq3} |f_{AB}(t_0)|=1\;\;\ \mbox{for}\;\
\mbox{some}\;\ 0<t_0<\infty\ee
 which can be interpreted as the signature of
perfect communication (or perfect state transfer) between $A$ and
$B$ in time $t_0$. The effect of the modulus in (\ref{eq3}) is that
the state at $B$, after transmission, will no longer be
$\ket{\psi}$, but will be of the form \be
\alpha\ket{0}+e^{i\phi}\beta\ket{1}. \ee The phase factor
$e^{i\phi}$ is not a problem because $\phi$ is independent of
$\alpha$ and $\beta$ and will thus be a known quantity for the
graph, which we can correct for with an appropriate phase gate (for
more details see for example \cite{8,9'',yung}).


The model we will consider is a distance-regular network consisting
of $N$ sites labeled by $\{1,2, ... ,N\}$ and diameter $d$. Then we
stratify the network with respect to a chosen reference site, say
$1$, and assume that the network  contains only the output site $N$
in its last stratum (i.e., $\ket{\phi_d}=\ket{N}$). At time $t=0$,
the qubit in the first (input) site of the network is prepared in
the state $\ket{\psi_{in}}$. We wish to transfer the state to the
$N$th (output) site of the network with unit efficiency after a
well-defined period of time. Although our qubits represent generic
two state systems, for the convenience of exposition we will use the
term spin as it provides a simple physical picture of the network.
The standard basis for an individual qubit is chosen to be
$\{|0\rangle=|\downarrow\rangle,\;\ |1\rangle=|\uparrow\rangle\}$,
and we shall assume that initially all spins point ``down" along a
prescribed $z$ axis; i.e., the network is in the state
$|\b{0}\rangle=|0_A00...00_B\rangle$. Then, we consider the dynamics
of the system to be governed by the quantum-mechanical Hamiltonian
\begin{equation}\label{H}
H_G =\frac{1}{2} \sum_{m=0}^dJ_{m}\sum_{_{(i,j)\in R_m}}H_{ij},
\end{equation}
with $H_{ij}$ as
\begin{equation}
H_{ij} = {\mathbf{\sigma}}_i\cdot {\mathbf{\sigma}}_j,
\end{equation}
where, ${\mathbf{\sigma}}_i$ is a vector with familiar Pauli
matrices $\sigma^x_i, \sigma^y_i$ and $\sigma^z_i$  as its
components acting on the one-site Hilbert space ${\mathcal{H}}_i$,
and $J_{m}$ is the coupling strength between the reference site
$1$ and all of the sites belonging to the $m$-th stratum with
respect to $1$.

The total spin of a quantum-mechanical system consisting of $N$
elementary spins $\vec{\sigma}_i$ on a one-dimensional lattice or
better called chain is given by:
\begin{equation}\label{S}
 \vec{\sigma}= \sum_{i=1}^N\vec{\sigma}_i.
\end{equation}
One can easily see that, the Hamiltonian (\ref{H}) commutes with
the total Spin operator (conservation). That is, since the total
$z$ component of the spin given by $\sigma^z_{tot}=\sum_{i\in
V}\sigma^z_i$ is conserved, i.e., $[\sigma^z_{tot},H_G]=0$, hence
the Hilbert space ${\mathcal{H}}_G$ decomposes into invariant
subspaces, each of which is a distinct eigenspace of the operator
$\sigma^z_{tot}$ (this property would be important to use its
symmetry to diagonalize the Hamiltonian in the well known Bethe
ansatz approach).

In order to consider perfect quantum state transfer, we write the
hamiltonian (\ref{H}) in terms of the adjacency matrices $A_i$,
$i=0,1,...,d$ of the underlying graph in order to use the techniques
introduced in section $2$ such as stratification and spectral
distribution associated with the graph. To do so, we recall that the
kets $\ket{i_1,i_2,...,i_N}$ with $i_1,...,i_N\in
\{\uparrow,\downarrow\}$ form an orthonormal basis for Hilbert space
${\mathcal{H}}_G$. Then, one can easily obtain
$$H_{ij}\ket{...\underbrace{\uparrow}_i...\underbrace{\uparrow}_j...}=\ket{...\underbrace{\uparrow}_i...\underbrace{\uparrow}_j...},$$
\begin{equation}\label{HK}
H_{ij}\ket{...\underbrace{\uparrow}_i...\underbrace{\downarrow}_j...}=-\ket{...\underbrace{\uparrow}_i...\underbrace{\downarrow}_j...}+2\ket{...\underbrace{\downarrow}_i...\underbrace{\uparrow}_j...}.
\end{equation}
where, we have used the facts that
$\sigma_z\ket{\uparrow}=\ket{\uparrow},\sigma_z\ket{\downarrow}=-\ket{\downarrow}$,
$\sigma_x\ket{\uparrow}=\ket{\downarrow},\sigma_x\ket{\downarrow}=\ket{\uparrow}$
and
$\sigma_y\ket{\uparrow}=i\ket{\downarrow},\sigma_y\ket{\downarrow}=-i\ket{\uparrow}$.
The equation (\ref{HK}) implies that the action of $H_{ij}$ on the
basis vectors is equivalent to the action of the operator
$2P_{ij}-I_N$, i.e., we have
\begin{equation}\label{HP}
H_{ij}=2P_{ij}-I_N,
\end{equation}
where, $P_{ij}$ denotes the permutation operator which permutes
$i$-th and $j$-th sites and $I_N$ is $N\times N$ identity matrix,
where $N$ is the number of vertices ($N:=|V|$). Now, let $\ket{l}$
denotes the vector state which its all components are $\uparrow$
except for $l$, i.e.,
$\ket{l}=\ket{\uparrow...\uparrow\underbrace{\downarrow}_l\uparrow...\uparrow}$.
Then, we have
$$\sum_{_{(i,j)\in R_m}}P_{ij}\ket{l}=\frac{1}{2}(\sum_{_{i\in \Gamma_m(j);i,j\neq l}}P_{ij}+2\sum_{_{i\in \Gamma_m(l)}}P_{il})\ket{l}=(\frac{ N\kappa_m}{2}-\kappa_m)\ket{l}+\sum_{j\in
\Gamma_m(l)}\ket{j},$$ which implies that
\begin{equation}\label{P}
\sum_{_{(i,j)\in R_m}}P_{ij}=(\frac{\kappa_m(N-2)}{2}I+A_m).
\end{equation}
Then, by using (\ref{HP}) and (\ref{P}), the hamiltonian in
(\ref{H}) can be written in terms of the adjacency matrices $A_i$,
$i=0,1,...,d$ as follows
\begin{equation}\label{HA}
H=\sum_{m=0}^dJ_m\sum_{_{(i,j)\in
R_m}}(2P_{ij}-I_N)=2\sum_{m=0}^dJ_mA_m+\frac{N-4}{2}\sum_{m=0}^dJ_m\kappa_m
I.
\end{equation}
As it has been shown in \cite{Kostak}, many known Hamiltonians
suitable for PST are basically associated with permutations and
can thus be obtained within the present unifying theoretical
framework. For the purpose of the perfect transfer of state, we
consider distance-regular graphs with $\kappa_d=|\Gamma_d(o)|=1$,
i.e., the last stratum of the graph contains only one site. Then,
we impose the constraints that the amplitudes
$\langle\phi_i|e^{-iHt}|\phi_0\rangle$ be zero for all
$i=0,1,...,d-1$ and
$\langle\phi_d|e^{-iHt}|\phi_0\rangle=e^{i\theta}$, where $\theta$
is an arbitrary phase. Therefore, these amplitudes must be
evaluated. To do so, we use the stratification and spectral
distribution associated with distance-regular graphs to write
$$\langle\phi_i|e^{-iHt}|\phi_0\rangle=e^{-\frac{i(N-4)t}{2}\sum_{m=0}^dJ_m\kappa_m}\langle\phi_i|e^{-2it\sum_{m=0}^dJ_mA_m}|\phi_0\rangle=$$
$$\frac{1}{\sqrt{\kappa_i}}e^{-\frac{i(N-4)t}{2}\sum_{m=0}^dJ_m\kappa_m}\langle\phi_0|A_ie^{-2it\sum_{m=0}^dJ_mP_m(A)}|\phi_0\rangle$$
Let the spectral distribution of the graph is
$\mu(x)=\sum_{k=0}^d\gamma_k\delta(x-x_k)$. Then,
$\langle\phi_i|e^{-iHt}|\phi_0\rangle=0$ implies that
$$\sum_{k=0}^d\gamma_kP_i(x_k)e^{-2it\sum_{m=0}^dJ_mP_m(x_k)}=0, \;\;\ i=0,1,...,d-1$$
Denoting $e^{-2it\sum_{m=0}^dJ_mP_m(x_k)}$ by $\eta_k$, the above
constraints are rewritten as follows
$$
\sum_{k=0}^dP_i(x_k)\eta_k\gamma_k=0,\;\;\ i=0,1,...,d-1,$$
\begin{equation}\label{Cons.}
\sum_{k=0}^dP_d(x_k)\eta_k\gamma_k=e^{i\theta}.
\end{equation}
As it was discussed previously, $P_i(x_k)$ are entries of the matrix
$\mathrm{P}$ (${\mathrm{P}}_{ki}=P_i(x_k)$) which is invertible,
i.e., the Eq.(\ref{Cons.}) can be written as
\begin{equation}\label{Cons.1}
\left(\begin{array}{c}
  \eta_0\gamma_0 \\
  \eta_1\gamma_1 \\
  \vdots \\
  \eta_d\gamma_d
\end{array}\right)=(P^t)^{-1}\left(\begin{array}{c}
  0 \\
  \vdots\\
  0 \\
  e^{i\theta}
\end{array}\right).
\end{equation}
The above equation implies that $\eta_k\gamma_k$ for $k=0,1,...,d$
are the same as the entries in the last column of the matrix
${(\mathrm{P}^t)}^{-1}=\frac{1}{v}\mathrm{Q^t}$ multiplied with
the phase $e^{i\theta}$, i.e., for the purpose of PST, the
following equations must be satisfied \be\label{result}
\eta_k\gamma_k=\gamma_ke^{-2it_0\sum_{m=0}^dJ_mP_m(x_k)}=\frac{e^{i\theta}}{v}{(\mathrm{Q^t})}_{kd}\;\
, \;\;\ \mbox{for} \;\ k=0,1,...,d .\ee In the following, we
investigate PST between antipodes of some distance-regular
networks such as cycle networks with even number of nodes and $d$
dimensional hypercube networks.
\subsection{Examples}
\textbf{1. Cycle graph $C_{2m}$}\\
A well known example of distance-regular networks, is the cycle
graph with $N$ vertices denoted by $C_N$ (see Fig. $1$ for even
$N=2m$). For the purpose of perfect transfer of state, we consider
the cycle graph with even number of vertices, since as it can be
seen from Fig. $1$, in this case the last stratum contains a
single state corresponding to the $m$-th vertex. From Figure $1$
it can  be also seen that, for even number of vertices $N=2m$, the
adjacency matrices are given by
\begin{equation}\label{adjcyce.}
A_0=I_{2m},\;\;\ A_i=S^i+S^{-i},\;\ i=1,2,...,m-1, \;\;\ A_m=S^m,
\end{equation}
where, $S$ is the $N\times N$ circulant matrix with period $N$ (
$S^N=I_N$) defined as follows
\begin{equation}
S=\left(\begin{array}{ccccc}
    0 & 1 & 0 & ... & 0 \\
    \vdots & \ddots & \ddots & \ddots & \vdots \\
    0 & \ldots & 0 & 1 & 0 \\
    0 & 0 & \ldots & 0 & 1 \\
    1 & 0 & \ldots & 0 & 0 \\
  \end{array}\right).
\end{equation}
By using (\ref{adjcyce.}), one can obtain the following recursion
relations for $C_{2m}$
\begin{equation}\label{adjcyce.0}
A_1A_i=A_{i-1}+A_{i+1},\;\ i=0,1,...,m-1; \;\ A_1A_m=A_{m-1}
\end{equation}
(the graph $C_{2m}$ consists of $m+1$ strata). By comparing
(\ref{adjcyce.0}) with three term recursion relations (\ref{P0}), we
obtain the intersection arrays for $C_{2m}$ as
\begin{equation}\label{intcye.}
\{b_0,...,b_{m-1};c_1,...,c_m\}=\{2,1,...,1,1;1,...,1,2\}.
\end{equation}Then, by using (\ref{omegal}), the
QD parameters are given by
\begin{equation}\label{QDcye.}
\alpha_i=0, \;\ i=0,1,...,m; \;\ \omega_1=\omega_m=2,\;\
\omega_i=1,\;\ i=2,...,m-1,
\end{equation}
By using the recursion relations (\ref{op}), one can show that \be
Q_0(x)=P_0(x)=1,\;\ Q_i(x)=P_i(x)=2T_i(x/2) ,\;\ i=1,..., m-1,\;\;\
Q_m(x)=2P_m(x)=2T_m(x/2)\ee where $T_i$'s are Chebyshev polynomials
of the first kind.

Then, the eigenvalues of the adjacency matrix $A\equiv A_1$ (roots
of $Q_{m+1}(x)=2T_{m+1}(x/2)$) are given by
$$x_i=\omega^i+\omega^{-i}=2\cos(2\pi i/N),\;\ i=0,1,...,m$$
with $\omega:=e^{2\pi i/N}$. Also, one can show that $\gamma_i$'s
(degeneracies of eigenvalues $x_i$) are given by \be\label{b}
\gamma_0=\gamma_m=1/2m,\;\;\ \gamma_i=1/m,\;\ i=1,2,...,m-1.\ee
Now, as regards the Eq. (\ref{push}), the matrix $P^t$ associated
with cycle graph $C_{2m}$ reads as \be
P^t=\left(\begin{array}{ccccc}
                                1 & 1 & \cdots & 1& 1 \\
                                2 & 2\cos(2\pi/N) & \cdots & 2\cos(2(m-1)\pi/N)& 2\omega^m \\
                                \vdots & \vdots & \vdots & \vdots & \vdots\\
                                2 & 2\cos(2(m-1)\pi/N) & \cdots & 2\cos((m-1)^2\pi/N)& 2\omega^{m(m-1)} \\
                                1 & \omega^m & \cdots & \cdots & \omega^{m^2} \\
                              \end{array}\right).\ee
One can see that $(P^t)^2=NI$, so the inverse of $P^t$ is given by
$(P^t)^{-1}=\frac{1}{N}P^t$. Therefore, by using (\ref{Cons.1})
and (\ref{b}), we obtain \be
\eta_i=e^{-it\sum_{l=0}^m2J_lT_l(\cos(2\pi
i/N))}=(-1)^ie^{i\theta} ,\;\;\ i=0,1,...,m\ee For instance, for
$N=4$, we obtain
$$\eta_0=e^{-it_0(J_0+2J_1+J_2)}=e^{i\theta},$$
$$\eta_1=e^{-it_0(J_0-J_2)}=-e^{i\theta},$$
\be \label{eqx} \eta_2=e^{-it_0(J_0-2J_1+J_2)}=e^{i\theta}\ee
which gives us the following equations
$$-t(J_0+2J_1+J_2)=\theta+2l\pi,$$
$$-t(J_0-J_2)=\theta+(2l'+1)\pi,$$
\be \label{eqx'}-t(J_0-2J_1+J_2)=\theta+2l''\pi.\ee For
$l=l'=l''=0$, one can obtain \be J_0=-\frac{2\theta+\pi}{4t_0},\;\
J_1=0,\;\; J_2=\frac{\pi}{4t_0},\ee whereas by choosing $l=l'=0,
l''=1$, the solution to (\ref{eqx'}) is given  by\be
J_0=-\frac{\theta+\pi}{2t_0},\;\ J_1=\frac{\pi}{4t_0},\;\;
J_2=0.\ee
 In the first case, the
time $t_0$ at which the state $\ket{\phi_0}=\ket{0}=\ket{1000}$ is
perfectly transferd to the vertex $\ket{\phi_2}=\ket{2}=\ket{0010}$
is given by \be t_0=-\frac{2\theta+\pi}{4J_0}=\frac{\pi}{4J_2},\ee
whereas in the latter case $t_0$ is given by \be
t_0=-\frac{\theta+\pi}{2J_0}=\frac{\pi}{4J_1}.\ee
\textbf{2. Hypercube network}\\
The hypercube of dimension $d$ (known also as binary Hamming
scheme $H(d,2)$) is a network with $N=2^d$ nodes, each of which
can be labeled by an $d$-bit binary string. Two nodes on the
hypercube described by bitstrings $\vec{x}$ and $\vec{y}$ are are
connected by an edge if $|\vec{x}- \vec{y}|=1$, where $|\vec{x}|$
is the Hamming weight of $\vec{x}$. In other words, if $\vec{x}$
and $\vec{y}$ differ by only a single bit flip, then the two
corresponding nodes on the network are connected. Thus, each of
$2^d$ nodes on the hypercube has degree $d$. For the hypercube
network with dimension $d$ we have $d+1$ strata with
\begin{equation}\label{cubevalenc}
\kappa_i=\frac{d!}{i!(d-i)!}\;\ , \;\ 0\leq i\leq d-1.
\end{equation}
The intersection numbers are given by
\begin{equation}
b_i=d-i,\;\;\ 0\leq i\leq d-1; \;\;\;\ c_i=i,\;\;\ 1\leq i\leq d.
\end{equation}
Furthermore, the adjacency matrices of this network are given by
\begin{equation}
A_i=\sum_{perm.}\underbrace{\sigma_x\otimes\sigma_x...\otimes\sigma_x}_{i}
\underbrace{\otimes I_2\otimes...\otimes I_2}_{n-i},\;\ i=0,1,...,n,
\end{equation}
where, the summation is taken over all possible nontrivial
permutations. In fact, the underlying network is the cartesian
product of $d$-tuples of complete network $K_2$. Also it can be
shown that, the idempotents $\{E_0,E_1,...,E_d\}$ are symmetric
product of $d$-tuples of corresponding idempotents of complete
network $K_2$. That is, we have
\begin{equation}
E_i=\sum_{perm.}\underbrace{E_-\otimes E_-...\otimes E_-}_{i}
\underbrace{\otimes E_+\otimes...\otimes E_+}_{d-i},\;\;\
i=0,1,...,d,
\end{equation}
where
\begin{equation}
E_{\pm}=\frac{1}{2}(I\pm\sigma_x).
\end{equation}
It has been shown that the eigenmatrices $P$ and $Q$ for the Hamming
scheme $H(d, 2)$ are the same, i.e., this scheme is self dual
\cite{ham}. Also, Delsarte \cite{dels} showed that the entries of
the eigenmatrix $P=Q$ for the Hamming scheme $H(d, 2)$ can be found
using the Krawtchouk polynomials as follows
\begin{equation}\label{ham'}
P_{il}=Q_{il}=K_l(i),
\end{equation}
where $K_l(x)$ are the Krawtchouk polynomials defined as
\begin{equation}
K_l(x)=\sum_{i=0}^{l}
 \left(\begin{array}{c}
   x \\
   i \\
       \end{array}\right)\left(\begin{array}{c}
   d-x \\
   l-i \\
       \end{array}\right)(-1)^i.
\end{equation}
Therefore, we have $((P^t)^{-1})_{il}=\frac{1}{2^d}
Q_{li}=\frac{1}{2^d}K_i(l)$.

The eigenvalues $x_l$ of the adjacency matrix $A\equiv A_1$ and
corresponding degeneracies $\gamma_l$ are given by
$$x_l=2l-d;$$
\be\label{ham1} \gamma_l=\frac{d!}{2^dl!(d-l)!},\;\;\
l=0,1,...,d.\ee By using (\ref{ham'}), we have
\be\eta_l=e^{-2it\sum_{m=0}^dJ_mK_m(l)},\;\ l=0,1,...,d.\ee Now, in
order to evaluate the time $t_0$ at which PST takes place, the
following equations must be satisfied
$$\eta_l\gamma_l=\frac{e^{i\theta}}{2^d}Q_{dl}=\frac{e^{i\theta}}{2^d}K_l(d), \;\ \forall \;\
l=0,1,...,d,$$ which are equivalent to
\be\frac{d!}{l!(d-l)!}e^{-2it_0\sum_{m=0}^dJ_mK_m(l)}=e^{i\theta}K_l(d)\;\
, \;\ \forall \;\ l=0,1,...,d.\ee For instance, in the case of $d=3$
(see Fig. 2), we must solve the following equations
$$e^{-2it_0(J_0+3J_1+3J_2+J_3)}=e^{i\theta},$$
$$e^{-2it_0(J_0+J_1-J_2-J_3)}=-e^{i\theta},$$
$$e^{-2it_0(J_0-J_1-J_2+J_3)}=e^{i\theta}$$
\be \label{eqH3} e^{-2it_0(J_0-3J_1+3J_2-J_3)}=-e^{i\theta}.\ee By
solving Eqs. (\ref{eqH3}) one can obtain the following solution
\be J_0=-\frac{2\theta+3\pi}{4t_0},\;\ J_1=-\frac{\pi
J_0}{2\theta+3\pi}=\frac{\pi}{4t_0},\;\ J_2=J_3=0;  \;\;\
\theta\neq -3\pi/2\ee that is the time $t_0$ at which PST takes
place is given by \be
t_0=-\frac{2\theta+3\pi}{4J_0}=\frac{\pi}{4J_1}.\ee

In the appendix we consider PST over antipodes of some important
finite distance-regular networks.
\section{Conclusion}
 By using spectral analysis techniques and algebraic combinatoric structures of
distance-regular graphs (as spin networks) such as stratification
introduced in \cite{obata, js} and Bose-Mesner algebra, a method
for finding a set of coupling constants in some particular spin
Hamiltonians associated with spin networks of distance-regular
type was given so that perfect state transfer between antipodes of
the networks can be achieved. As examples, the cycle networks with
even number
of vertices and $d$-dimensional hypercube networks were considered.\\

\newpage
 \vspace{1cm}\setcounter{section}{0}
 \setcounter{equation}{0}
 \renewcommand{\theequation}{A-\roman{equation}}
  {\Large{Appendix}}\\
In this appendix we consider some important finite distance-regular
networks such that their last stratum contains only one node. Then
by using the prescription of section 3, we
investigate PST over antipodes  of these networks.\\
\textbf{1. Icosahedron} \cite{14}\\
Intersection array:
$$\{b_0,b_1,b_2;c_1,c_2,c_3\}=\{5,2,1;1,2,5\}.$$ Size of strata
and QD parameters:
$$\kappa_0=1,\;\ \kappa\equiv \kappa_1=5,\;\ \kappa_2=5,\;\ \kappa_3=1,$$
$$\alpha_0=0,\;\ \alpha_1=\alpha_2=2,\;\ \alpha_3=0; \;\;\ \omega_1=5,\;\ \omega_2=4,\;\ \omega_3=5.$$
Polynomials $P_i(x)$:
$$P_0=1,\;\ P_1(x)=x,\;\
P_2(x)=\frac{1}{2}(x^2-2x-5),\;\
P_3(x)=\frac{1}{10}(x^3-4x^2-5x+10).$$
Stieltjes function:
$$G_{\mu}(x)=\frac{x^3-4x^2-5x+10}{x^4-4x^3-10x^2+20x+25}.$$
Spectral distribution (
$\mu(x)=\sum_{l=0}^d\gamma_l\delta(x-x_l)$):
$$\mu(x)=\frac{1}{12}\{5\delta(x+1)+\delta(x-5)+3\delta(x-\sqrt{5})+3\delta(x+\sqrt{5})\}.$$
 Now, one can obtain the matrix $P^t$ and its inverse.
Then by solving the equations (\ref{result}), the solution is
obtained as follows \be J_0=-\frac{6\theta+7\pi}{12t_0},\;\
J_1=-\frac{(5-3\sqrt{5})\pi}{60t_0},\;\
J_2=-\frac{(5+3\sqrt{5})\pi}{60t_0},\;\ J_3=\frac{5\pi}{12t_0}.\ee
Then, the time $t_0$ at which PST takes place is given by \be
t_0=-\frac{2\theta+\pi}{4J_0}=\frac{\pi}{4J_3}.\ee
\textbf{2. Desargues} \cite{14}\\
Intersection array:
$$\{b_0,b_1,b_2,b_3,b_4;c_1,c_2,c_3,c_4,c_5\}=\{3,2,2,1,1;1,1,2,2,3\}.$$
Size of strata and QD parameters:
$$\kappa_0=1,\;\ \kappa\equiv \kappa_1=3,\;\ \kappa_2=6,\;\ \kappa_3=6,\;\ \kappa_4=3,\;\ \kappa_5=1$$
$$\alpha_i=0,\;\ i=0,1,...,5; \;\;\ \omega_1=3,\;\ \omega_2=2,\;\ \omega_3=4,\;\ \omega_4=2,\;\ \omega_5=3.$$
Polynomials $P_i(x)$:
$$\hspace{-1.7cm}P_0=1,\;\ P_1(x)=x,\;\ P_2(x)=x^2-3,\;\ P_3(x)=\frac{1}{2}(x^3-5x),\;\ P_4(x)=\frac{1}{4}(x^4-9x^2+12),\;\ P_5(x)=\frac{1}{12}(x^5-11x^3+22x).$$
Stieltjes function:
$$G_{\mu}(x)=\frac{x^5-11x^3+22x}{x^6-14x^4+49x^2-36}.$$
Spectral distribution:
$$\mu(x)=\frac{1}{20}\{5\delta(x+1)+5\delta(x-1)+4\delta(x+2)+4\delta(x-2)+\delta(x+3)+\delta(x-3)\}.$$ The solution to Eq.(\ref{result}) is
given by:
$$J_0=-\frac{30\theta+51\pi}{60t_0},\;\;\ J_1=\frac{\pi}{10t_0},\;\ J_2=-\frac{4\pi}{15t_0},\;\ J_3=0,\;\  J_4=\frac{\pi}{15t_0},\;\ J_5=\frac{\pi}{4t_0}.$$
\textbf{3. Dodecahedron} \cite{14}\\
Intersection array:
$$\{b_0,b_1,b_2,b_3,b_4;c_1,c_2,c_3,c_4,c_5\}=\{3,2,1,1,1;1,1,1,2,3\}.$$
Size of strata and QD parameters:
$$\kappa_0=1,\;\ \kappa\equiv \kappa_1=3,\;\ \kappa_2=6,\;\ \kappa_3=6,\;\ \kappa_4=3,\;\ \kappa_5=1$$
$$\alpha_0=\alpha_1=0,\;\ \alpha_2=\alpha_3=1,\;\ \alpha_4=\alpha_5=0; \;\;\ \omega_1=3,\;\ \omega_2=2,\;\ \omega_3=1,\;\ \omega_4=2,\;\ \omega_5=3.$$
Polynomials $P_i(x)$:
$$P_0=1, \;\ P_1(x)=x, \;\ P_2(x)=x^2-3,\;\ P_3(x)=x^3-5x-x^2+3,\;\ P_4(x)=\frac{1}{2}(x^4-5x^2-2x^3+8x),$$
$$ P_5(x)=\frac{1}{6}(x^5-7x^3-2x^4+10x^2+10x-6).$$ Stieltjes
function:
$$G_{\mu}(x)=\frac{x^5-7x^3-2x^4+10x^2+10x-6}{x^6-10x^4-2x^5+16x^3+25x^2-30x}.$$
Spectral distribution:
$$\mu(x)=\frac{1}{20}\{4\delta(x)+5\delta(x-1)+4\delta(x+2)+\delta(x-3)+3\delta(x-\sqrt{5})+3\delta(x+\sqrt{5})\}.$$ The solution to Eq.(\ref{result}) is
given by:
$$J_0=-\frac{\theta+2\pi}{2t_0},\;\;\ J_1=\frac{(2+3\sqrt{5})\pi}{60t_0},\;\;\ J_2=-\frac{17\pi}{60t_0},\;\;\ J_3=\frac{\pi}{60t_0},\;\;\ J_4=\frac{(2-3\sqrt{5})\pi}{60t_0}.$$
\textbf{4. Taylor($P(13)$)} \cite{vandam}\\
Intersection array:\\
$\{b_0,b_1,b_2;c_1,c_2,c_3\}=\{13,6,1;1,6,13\}$. Size of strata
and QD parameters:
$$\kappa_0=1,\;\ \kappa\equiv \kappa_1=13,\;\ \kappa_2=13,\;\ \kappa_3=1,$$
$$\alpha_0=0,\;\ \alpha_1=\alpha_2=6,\;\ \alpha_3=0; \;\;\ \omega_1=13,\;\ \omega_2=36,\;\ \omega_3=13.$$
Polynomials $P_i(x)$:
$$\hspace{-1.7cm}P_0=1,\;\ P_1(x)=x,\;\ P_2(x)=\frac{1}{6}(x^2-6x-13),\;\ P_3(x)=\frac{1}{78}(x^3-12x^2-13x+78).$$
Stieltjes function:
$$G_{\mu}(x)=\frac{x^3-12x^2-13x+78}{x^4-12x^3-26x^2+156x+169}.$$
Spectral distribution:
$$\mu(x)=\frac{1}{28}\{13\delta(x+1)+\delta(x-13)+7\delta(x-\sqrt{13})+7\delta(x+\sqrt{13})\}.$$ The solution to Eq.(\ref{result}) is
given by:
$$J_0=-\frac{14\theta+15\pi}{28t_0},\;\;\ J_1=-\frac{(13-7\sqrt{13})\pi}{364t_0},\;\;\ J_2=-\frac{(13+7\sqrt{13})\pi}{364t_0},\;\;\ J_3=\frac{13\pi}{28t_0}.$$
\textbf{5. Taylor($GQ(2,2)$)} \cite{vandam}\\
Intersection
array:$$\{b_0,b_1,b_2;c_1,c_2,c_3\}=\{15,8,1;1,8,15\}.$$ Size of
strata and QD parameters:
$$\kappa_0=1,\;\ \kappa\equiv \kappa_1=15,\;\ \kappa_2=15,\;\ \kappa_3=1,$$
$$\alpha_0=0,\;\ \alpha_1=\alpha_2=6,\;\ \alpha_3=0; \;\;\ \omega_1=15,\;\ \omega_2=64,\;\ \omega_3=15.$$
Polynomials $P_i(x)$:
$$\hspace{-1.7cm}P_0=1,\;\ P_1(x)=x,\;\ P_2(x)=\frac{1}{8}(x^2-6x-15),\;\ P_3(x)=\frac{1}{120}(x^3-12x^2-43x+90).$$
Stieltjes function:
$$G_{\mu}(x)=\frac{x^3-12x^2-43x+90}{x^4-12x^3-58x^2+180x+225}.$$
Spectral distribution:
$$\mu(x)=\frac{1}{32}\{15\delta(x+1)+10\delta(x-3)+6\delta(x+5)+\delta(x-15)\}.$$ The solution to Eq.(\ref{result}) is
given by:
$$J_0=-\frac{16\theta+15\pi}{32t_0},\;\;\ J_1=\frac{\pi}{32t_0},\;\;\ J_2=-\frac{3\pi}{32t_0},\;\;\ J_3=\frac{13\pi}{32t_0}.$$
\textbf{6. Taylor($T(6)$)} \cite{vandam}\\
Intersection array:
$$\{b_0,b_1,b_2;c_1,c_2,c_3\}=\{15,6,1;1,6,15\}.$$ Size of strata
and QD parameters:
$$\kappa_0=1,\;\ \kappa\equiv \kappa_1=15,\;\ \kappa_2=15,\;\ \kappa_3=1,$$
$$\alpha_0=0,\;\ \alpha_1=\alpha_2=8,\;\ \alpha_3=0; \;\;\ \omega_1=15,\;\ \omega_2=36,\;\ \omega_3=15.$$
Polynomials $P_i(x)$:
$$\hspace{-1.7cm}P_0=1,\;\ P_1(x)=x,\;\ P_2(x)=\frac{1}{6}(x^2-8x-15),\;\ P_3(x)=\frac{1}{90}(x^3-16x^2+13x+120).$$
Stieltjes function:
$$G_{\mu}(x)=\frac{x^3-16x^2+13x+120}{x^4-16x^3-2x^2+240x+225}.$$
Spectral distribution:
$$\mu(x)=\frac{1}{32}\{15\delta(x+1)+10\delta(x+3)+6\delta(x-5)+\delta(x-15)\}.$$The solution to Eq.(\ref{result}) is
given by:
$$J_0=-\frac{16\theta+15\pi}{32t_0},\;\;\ J_1=-\frac{3\pi}{32t_0},\;\;\ J_2=\frac{\pi}{32t_0},\;\;\ J_3=\frac{13\pi}{32t_0}.$$
\textbf{7. Wells} \cite{7'}\\
Intersection array:
$$\{b_0,b_1,b_2,b_3;c_1,c_2,c_3,c_4\}=\{5,4,1,1;1,1,4,5\}.$$
Size of strata and QD parameters:
$$\kappa_0=1,\;\ \kappa\equiv \kappa_1=5,\;\ \kappa_2=20,\;\ \kappa_3=5,\;\ \kappa_4=1, $$
$$\alpha_0=\alpha_1=0,\;\ \alpha_2=3,\;\ \alpha_3=\alpha_4=0; \;\;\ \omega_1=5,\;\ \omega_2=\omega_3=4,\;\ \omega_4=5.$$
Polynomials $P_i(x)$:
$$\hspace{-0.5cm}P_0=1,\;\ P_1(x)=x,\;\ P_2(x)=x^2-5,\;\ P_3(x)=\frac{1}{4}(x^3-9x-3x^2+15),\;\ P_4(x)=\frac{1}{20}(x^4-13x^2-3x^3+15x+20).$$
Stieltjes function:
$$G_{\mu}(x)=\frac{x^4-13x^2-3x^3+15x+20}{x^5-18x^3-3x^4+30x^2+65x-75}.$$
Spectral distribution:
$$\mu(x)=\frac{1}{32}\{10\delta(x-1)+5\delta(x+3)+\delta(x-5)+
8\delta(x+\sqrt{5})+8\delta(x-\sqrt{5})\}.$$ The solution to
Eq.(\ref{result}) is given by:
$$J_0=-\frac{16 \theta+23\pi}{32t_0},\;\;\ J_1=\frac{(5-8\sqrt{5})\pi}{160t_0},\;\;\ J_2=-\frac{3\pi}{32t_0},\;\;\ J_3=\frac{(5+8\sqrt{5})\pi}{160t_0},\;\;\ J_4=\frac{9\pi}{32t_0}.$$
\textbf{8. Hadamard network} \cite{9'}\\
Intersection array:
$$\{b_0,b_1,b_2,b_3;c_1,c_2,c_3,c_4\}=\{8,7,4,1;1,4,7,8\}.$$ Size of strata and
QD parameters:
$$\kappa_0=1,\;\ \kappa\equiv \kappa_1=8,\;\ \kappa_2=14,\;\ \kappa_3=8,\;\ \kappa_4=1,$$
$$\alpha_i=0,\;\ i=0,1,...,4; \;\;\ \omega_1=8,\;\ \omega_2=28,\;\ \omega_3=28,\;\ \omega_4=8.$$
Polynomials $P_i(x)$:
$$\hspace{-1.7cm}P_0=1,\;\ P_1(x)=x,\;\ P_2(x)=\frac{1}{4}(x^2-8),\;\ P_3(x)=\frac{1}{28}(x^3-36x),\;\ P_4(x)=\frac{1}{224}(x^4-64x^2+224).$$
Stieltjes function:
$$G_{\mu}(x)=\frac{x^4-64x^2+224}{x^5-72x^3+512x}.$$
Spectral distribution:
$$\mu(x)=\frac{1}{32}\{14\delta(x)+\delta(x-8)+\delta(x+8)+8\delta(x-2\sqrt{2})+8\delta(x+2\sqrt{2})\}.$$ The solution to Eq.(\ref{result}) is
given by:
$$J_0=-\frac{16\theta+19\pi}{32t_0},\;\;\ J_1=\frac{(1+2\sqrt{2})\pi}{32t_0},\;\;\
J_2=-\frac{3\pi}{32t_0},\;\;\
J_3=\frac{(1-2\sqrt{2})\pi}{32t_0},\;\;\
J_4=\frac{13\pi}{32t_0}.$$
\textbf{9. Taylor($P(17)$)} \cite{vandam}\\
Intersection array:\\
$\{b_0,b_1,b_2;c_1,c_2,c_3\}=\{17,8,1;1,8,17\}$. Size of strata
and QD parameters:
$$\kappa_0=1,\;\ \kappa\equiv \kappa_1=17,\;\ \kappa_2=17,\;\ \kappa_3=1,$$
$$\alpha_0=0,\;\ \alpha_1=\alpha_2=8,\;\ \alpha_3=0; \;\;\ \omega_1=17,\;\ \omega_2=64,\;\ \omega_3=17.$$
Polynomials $P_i(x)$:
$$\hspace{-1.7cm}P_0=1,\;\ P_1(x)=x,\;\ P_2(x)=\frac{1}{8}(x^2-8x-17),\;\ P_3(x)=\frac{1}{136}(x^3-16x^2-17x+136).$$
Stieltjes function:
$$G_{\mu}(x)=\frac{x^3-16x^2-17x+136}{x^4-16x^3-34x^2+272x+289}.$$
Spectral distribution:
$$\mu(x)=\frac{1}{36}\{17\delta(x+1)+\delta(x-17)+9\delta(x-\sqrt{17})+9\delta(x+\sqrt{17})\}.$$The solution to Eq.(\ref{result}) is
given by:
$$J_0=-\frac{18 \theta+19\pi}{36t_0},\;\;\ J_1=-\frac{(17-9\sqrt{17})\pi}{612t_0},\;\;\ J_2=-\frac{(17+9\sqrt{17})\pi}{612t_0},\;\;\ J_3=\frac{17\pi}{36t_0}.$$
\textbf{10. Hadamard network} \cite{vandam}\\
Intersection array:
$$\{b_0,b_1,b_2,b_3;c_1,c_2,c_3,c_4\}=\{12,11,6,1;1,6,11,12\}.$$ Size of strata and
QD parameters:
$$\kappa_0=1,\;\ \kappa\equiv \kappa_1=12,\;\ \kappa_2=22,\;\ \kappa_3=12,\;\ \kappa_4=1,$$
$$\alpha_i=0,\;\ i=0,1,...,4; \;\;\ \omega_1=12,\;\ \omega_2=\omega_3=66,\;\ \omega_4=12.$$
Polynomials $P_i(x)$:
$$\hspace{-1.7cm}P_0=1,\;\ P_1(x)=x,\;\ P_2(x)=\frac{1}{6}(x^2-12),\;\ P_3(x)=\frac{1}{66}(x^3-78x),\;\ P_4(x)=\frac{1}{792}(x^4-144x^2+792).$$
Stieltjes function:
$$G_{\mu}(x)=\frac{x^4-144x^2+792}{x^5-156x^3+1728x}.$$Spectral distribution:
$$\mu(x)=\frac{1}{48}\{22\delta(x)+\delta(x+12)+\delta(x-12)+12\delta(x-2\sqrt{3})+12\delta(x+2\sqrt{3})\}.$$ The solution to Eq.(\ref{result}) is
given by:
$$J_0=-\frac{24\theta+27\pi}{48t_0},\;\;\ J_1=-\frac{(1-2\sqrt{3})\pi}{48t_0},\;\;\
J_2=-\frac{\pi}{16t_0},\;\;\
J_3=-\frac{(1+2\sqrt{3})\pi}{48t_0},\;\;\
J_4=\frac{21\pi}{48t_0}.$$
\textbf{11. Taylor($SRG(25,12)$)} \cite{vandam}\\
Intersection
array:$$\{b_0,b_1,b_2;c_1,c_2,c_3\}=\{25,12,1;1,12,25\}.$$ Size of
strata and QD parameters:
$$\kappa_0=1,\;\ \kappa\equiv \kappa_1=25,\;\ \kappa_2=25,\;\ \kappa_3=1,$$
$$\alpha_0=0,\;\ \alpha_1=\alpha_2=12,\;\ \alpha_3=0; \;\;\ \omega_1=25,\;\ \omega_2=144,\;\ \omega_3=25.$$
Polynomials $P_i(x)$:
$$\hspace{-1.7cm}P_0=1,\;\ P_1(x)=x,\;\ P_2(x)=\frac{1}{12}(x^2-12x-25),\;\ P_3(x)=\frac{1}{300}(x^3-24x^2-25x+300).$$
Stieltjes function:
$$G_{\mu}(x)=\frac{x^3-24x^2-25x+300}{x^4-24x^3-50x^2+600x+625}.$$Spectral distribution:
$$\mu(x)=\frac{1}{52}\{25\delta(x+1)+13\delta(x-5)+13\delta(x+5)+\delta(x-25)\}.$$ The solution to Eq.(\ref{result}) is
given by:
$$J_0=-\frac{26\theta+27\pi}{52t_0},\;\;\ J_1=\frac{2\pi}{65t_0},\;\;\
J_2=-\frac{9\pi}{130t_0},\;\;\ J_3=\frac{25\pi}{52t_0}.$$
\textbf{12. Gosset, Tayl(Schl\"{a}fli)} \cite{vandam}\\
Intersection
array:$$\{b_0,b_1,b_2;c_1,c_2,c_3\}=\{27,10,1;1,10,27\}.$$ Size of
strata and QD parameters:
$$\kappa_0=1,\;\ \kappa\equiv \kappa_1=27,\;\ \kappa_2=27,\;\ \kappa_3=1,$$
$$\alpha_0=0,\;\ \alpha_1=\alpha_2=16,\;\ \alpha_3=0; \;\;\ \omega_1=27,\;\ \omega_2=100,\;\ \omega_3=27.$$
Polynomials $P_i(x)$:
$$\hspace{-1.7cm}P_0=1,\;\ P_1(x)=x,\;\ P_2(x)=\frac{1}{10}(x^2-16x-27),\;\ P_3(x)=\frac{1}{270}(x^3-32x^2+129x+432).$$
Stieltjes function:
$$G_{\mu}(x)=\frac{x^3-32x^2+129x+432}{x^4-32x^3+102x^2+864x+729}.$$Spectral distribution:
$$\mu(x)=\frac{1}{56}\{27\delta(x+1)+21\delta(x+3)+7\delta(x-9)+\delta(x-27)\}.$$ The solution to Eq.(\ref{result}) is
given by:
$$J_0=-\frac{14\theta+11\pi}{28t_0},\;\;\ J_1=-\frac{5\pi}{84t_0},\;\;\
J_2=\frac{\pi}{42t_0},\;\;\ J_3=\frac{5\pi}{14t_0}.$$
\textbf{13. Taylor(Co-Schl\"{a}fli))} \cite{vandam}\\
Intersection
array:$$\{b_0,b_1,b_2;c_1,c_2,c_3\}=\{27,16,1;1,16,27\}.$$ Size of
strata and QD parameters:
$$\kappa_0=1,\;\ \kappa\equiv \kappa_1=27,\;\ \kappa_2=27,\;\ \kappa_3=1,$$
$$\alpha_0=0,\;\ \alpha_1=\alpha_2=10,\;\ \alpha_3=0; \;\;\ \omega_1=27,\;\ \omega_2=256,\;\ \omega_3=27.$$
Polynomials $P_i(x)$:
$$\hspace{-1.7cm}P_0=1,\;\ P_1(x)=x,\;\ P_2(x)=\frac{1}{16}(x^2-10x-27),\;\ P_3(x)=\frac{1}{432}(x^3-20x^2-183x+270).$$
Stieltjes function:
$$G_{\mu}(x)=\frac{x^3-20x^2-183x+270}{x^4-20x^3-210x^2+540x+729}.$$ Spectral distribution:
$$\mu(x)=\frac{1}{56}\{27\delta(x+1)+21\delta(x-3)+7\delta(x+9)+\delta(x-27)\}.$$The solution to Eq.(\ref{result}) is
given by:
$$J_0=-\frac{14\theta+11\pi}{28t_0},\;\;\
J_1=\frac{\pi}{42t_0},\;\;\ J_2=-\frac{5\pi}{84t_0},\;\;\
J_3=\frac{5\pi}{14t_0}.$$
\textbf{14. Taylor($SRG(29,14)$)} \cite{vandam}\\
Intersection array:
$$\{b_0,b_1,b_2;c_1,c_2,c_3\}=\{29,14,1;1,14,29\}.$$ Size of strata and
QD parameters:
$$\kappa_0=1,\;\ \kappa\equiv \kappa_1=29,\;\ \kappa_2=29,\;\ \kappa_3=1,$$
$$\alpha_0=0,\;\ \alpha_1=\alpha_2=14,\;\ \alpha_3=0; \;\;\ \omega_1=29,\;\ \omega_2=196,\;\ \omega_3=29.$$
Polynomials $P_i(x)$:
$$\hspace{-1.7cm}P_0=1,\;\ P_1(x)=x,\;\ P_2(x)=\frac{1}{14}(x^2-14x-29),\;\ P_3(x)=\frac{1}{406}(x^3-28x^2-29x+406).$$
Stieltjes function:
$$G_{\mu}(x)=\frac{x^3-28x^2-29x+406}{x^4-28x^3-58x^2+812x+841}.$$Spectral distribution:
$$\mu(x)=\frac{1}{60}\{29 \delta(x+1)+\delta(x-29)+15 \delta(x-\sqrt{29})+15 \delta(x+\sqrt{29})\}.$$ The solution to Eq.(\ref{result}) is
given by:
$$J_0=-\frac{30\theta+31\pi}{60t_0},\;\;\
J_1=-\frac{(29-15\sqrt{29})\pi}{1740t_0},\;\;\
J_2=-\frac{(29+15\sqrt{29})\pi}{1740t_0},\;\;\
J_3=\frac{29\pi}{60t_0}.$$
\textbf{15. Doubled odd(4)} \cite{vandam}\\
Intersection array:
$$\{b_0,b_1,b_2,b_3,b_4,b_5,b_6;c_1,c_2,c_3,c_4,c_5,c_6,c_7\}=\{4,3,3,2,2,1,1;1,1,2,2,3,3,4\}.$$ Size of strata and
QD parameters:
$$\kappa_0=1,\;\ \kappa\equiv \kappa_1=4,\;\ \kappa_2=12,\;\ \kappa_3=18,\;\ \kappa_4=18,\;\ \kappa_5=12,\;\ \kappa_6=4,\;\ \kappa_7=1,$$
$$\alpha_i=0,\;\ i=0,1,...,7; \;\;\ \omega_1=4,\;\ \omega_2=3,\;\ \omega_3=6,\;\ \omega_4=4,\;\ \omega_5=6,\;\ \omega_6=3,\;\ \omega_7=4.$$
Polynomials $P_i(x)$:
$$\hspace{-1.7cm}P_0=1,\;\ P_1(x)=x,\;\ P_2(x)=x^2-4,\;\ P_3(x)=\frac{1}{2}(x^3-7x),\;\ P_4(x)=\frac{1}{4}(x^4-13x^2+24),$$
$$\hspace{-1cm}P_5(x)=\frac{1}{12}(x^5-17x^3+52x),\;\ P_6(x)=\frac{1}{36}(x^6-23x^4+130x^2-144),\;\ P_7(x)=\frac{1}{144}(x^7-26x^5+181x^3-300x).$$
Stieltjes function:
$$G_{\mu}(x)=\frac{x^7-26x^5+181x^3-300x}{x^8-30x^6+273x^4-820x^2+576}.$$
Spectral distribution:
$$\mu(x)=\frac{1}{70}\{14[\delta(x-1)+\delta(x+1)+\delta(x-2)+\delta(x+2)]+6[\delta(x-3)+\delta(x+3)]+\delta(x-4)+\delta(x+4)\}.$$ The solution to Eq.(\ref{result}) is
given by:
$$\hspace{-0.7cm}J_0=-\frac{70\theta+151\pi}{140t_0},\;\
J_1=0,\;\;\ J_2=-\frac{56\pi}{245t_0},\;\ J_3=0,\;\
J_4=\frac{4\pi}{105t_0},\;\ J_5=-\frac{\pi}{70t_0},\;\
J_6=-\frac{\pi}{35t_0},\;\ J_7=\frac{\pi}{4t_0}.$$
\textbf{16. J(8,4)} \cite{vandam}\\
Intersection array:
$$\{b_0,b_1,b_2,b_3;c_1,c_2,c_3,c_4\}=\{16,9,4,1;1,4,9,16\}.$$
Size of strata and QD parameters:
$$\kappa_0=1,\;\ \kappa\equiv \kappa_1=16,\;\ \kappa_2=36,\;\ \kappa_3=16,\;\ \kappa_4=1,$$
$$\alpha_0=0,\;\ \alpha_1=6,\;\ \alpha_2=8,\;\ \alpha_3=6,\;\ \alpha_4=0; \;\;\ \omega_1=16,\;\ \omega_2=36,\;\ \omega_3=36,\;\ \omega_4=16.$$
Polynomials $P_i(x)$:
$$P_0=1,\;\ P_1(x)=x,\;\ P_2(x)=\frac{1}{4}(x^2-6x-16),\;\ P_3(x)=\frac{1}{36}(x^3-14x^2-4x+128),$$
$$ P_4(x)=\frac{1}{576}(x^4-20x^3+44x^2+368x-192).$$ Stieltjes
function:
$$G_{\mu}(x)=\frac{x^4-20x^3+44x^2+368x-192}{x^5-20x^4+28x^3+592x^2-128x-2048}.$$
Spectral distribution:
$$\mu(x)=\frac{1}{70}\{\delta(x-16)+7 \delta(x-8)+14 \delta(x+4)+28 \delta(x+2)+20 \delta(x-2)\}.$$ The solution to Eq.(\ref{result}) is
given by:
$$J_0=-\frac{70\theta+199\pi}{140t_0},\;\;\
J_1=\frac{\pi}{35t_0},\;\;\ J_2=\frac{13\pi}{210t_0},\;\;\
J_3=-\frac{\pi}{14t_0},\;\ J_4=-\frac{17\pi}{140t_0}.$$

\newpage
{\bf Figure Captions}

{\bf Figure 1:} Denotes the cycle network $C_{2m}$, where the $m+1$
vertical dashed lines show the $m+1$ strata of the network.

{\bf Figure 2:} Shows the cube or Hamming scheme $H(3,2)$ with
vertex set $V=\{(ijk): i,j,k=0,1\}$, where the vertical dashed lines
denote the four strata of the cube.
\end{document}